\documentstyle[version2,aps]{revtex}   
\begin{document}
\begin{title}
  Doping Dependence of the Chemical Potential \\
  in Cuprate High-$T_{\rm c}$ Superconductors I:
  La$_{2-x}$Sr$_x$CuO$_4$
\end{title}
\draft
\author{G. Rietveld$^{a,b}$, M. Glastra$^{a}$, D. van der Marel
$^{a,c}$}
\begin{instit}
  Delft University of Technology, Department of Applied Physics,\hfill\\
  Lorentzweg 1, 2628 CJ Delft, The Netherlands.$^{a}$\hfill\\
  NMi-Van Swinden Laboratorium, Department of Electrical
  Standards,\hfill\\
  P.O. Box 654, 2600 AR\ \ Delft, The Netherlands.$^{b}$\hfill\\
  Materials Science Centre, Solid State Physics Laboratory,
  University of Groningen,\hfill\\
  Nijenborgh 4, 9747 AG Groningen, The Netherlands$^c$
\end{instit}
\date{\today}
\begin{abstract}
  A systematic study is performed of the doping dependence of the
  chemical potential $\mu$ in La$_{2-x}$Sr$_x$CuO$_4$ as a function
  of Sr content, using a well-characterized series of pellets and
  thin films.
  The measured shift of the chemical potential, as deduced from the
  changes in the photoelectron spectra, is compared with present
  models for the doping behaviour of $\mu$ in high-$T_{\rm c}$ materials.
  The results obtained can be best described assuming $\mu$ shifts
  due to the doping of a rigid narrow band.
\end{abstract}

\pacs{{\em Keywords:
  La$_{2-x}$Sr$_x$CuO$_4$, XPS, thin films,
  chemical potential/Fermi level, metal-insulator transition}}

\widetext
\section{Introduction}

One of the major questions related to the microscopic mechanism of
  superconductivity in the cuprate high-$T_{\rm c}$ materials concerns
  the problem of the proximity of the superconducting compounds to
  a correlation induced metal insulator transition.
  In clarifying such questions, photoelectron spectroscopy has played
  a large role.
  With this technique no density of states is found near the Fermi
  energy $E_{\rm F}$ for the parent insulating materials, whereas for the
  doped metallic samples a clear Fermi level is present in the
  measurements \cite{Shen88,Imer89}, proving the fermionic character
  of the states near $E_{\rm F}$.
  Another important result obtained with photoelectron spectroscopy
  is that angle resolved measurements indicate a Fermi surface that
  agrees well in size and shape with predictions of band structure
  calculations \cite{Manzke89,Olson89,Takahashi88a,King93,Anderson93}.
  This agreement is remarkable, because in these calculations the
  correlations between the electrons is generally neglected.
  It is by now widely accepted that such correlations do play a large
  role in the high-$T_{\rm c}$ materials, most prominently visible in the
  fact that the parent compounds are insulators whereas bandstructure
  calculations predict them to be metals.

In this study, we will use X-ray photoelectron spectroscopy for studying
  the behaviour of the chemical potential $\mu$ as a function of doping
  in the high-$T_{\rm c}$ cuprates, especially in the neighbourhood of
  the metal-insulator transition.
  This behaviour of $\mu$ can be directly related to the low-energy
  electronic structure of these materials.
  Concerning the question of how to describe the behaviour of $\mu$
  as a function of doping, two possible descriptions are given in the
  literature up to now, schematically depicted in
  Fig.~\ref{fig:mux.models}.

The first model assumes that the states created near the Fermi level upon
  doping are induced by impurities. In the undoped material the chemical
  potential lies somewhere in the middle of
  the charge-transfer gap. Doping of the materials is achieved by
  the addition of impurities. These impurities create states near
  $\mu$ that are either directly related to the impurities or more
  indirectly generated by a impurity potential that pushes states
  from the valence band up to the Fermi level. The consequence of
  doping, {\it i.e.} the addition of impurities, thus is a filling of the
  gap with states. The chemical potential will be more or less constant,
  near the middle of the gap. The increase of the density of states near
  $\mu$ for finite doping is caused by the increase of the impurity induced
  density of states in the gap [Fig.~\ref{fig:mux.models}b].
The second model assigns the increase of density of states near $E_{\rm F}$
  to a shift of the chemical potential. Again, in the undoped case the
  chemical potential is situated near the middle of the gap. If the material
  is doped with holes, the chemical potential rapidly moves to the top of the
  valence band. At still higher doping, $\mu$ shifts further into the valence
  band.
  The increase in density of states near $\mu$ thus (at least partly) arises
  from a change in position of the chemical potential, with possible extra
  contributions of changes in electronic structure \cite{Eskes91}.
  The behaviour of $\mu$ according to this model is depicted
  in Fig.~\ref{fig:mux.models}c.
In the following we will refer to these two models as the `impurity
  model' and `semiconductor model' respectively. The latter name is
  because of the similarity of the behaviour of $\mu$ in this model
  to doping of a simple semiconductor.

Experimentally, the doping dependence of the chemical potential in a
material can
  be measured with photoelectron spectroscopy (PES) by determining
  the binding energies of all core levels and the valence band. In a
  photoelectron spectrum, the binding energies are measured with respect
  to the common Fermi level of the sample and electron analyzer.
  If the reference level $\mu$ changes upon doping of the material,
  this becomes visible as a collective shift of {\em all\/} core
  levels and the valence band.

Compared to the large amount of PES data in the literature, only a
  few serious studies devoted to the doping behaviour of $\mu$ have
  appeared.
  The main evidence in favour of the `impurity model' was put forward
  by Allen {\it et al.} \cite{Allen90}. They compared the valence band
  spectra of undoped and optimally doped Nd$_{2-x}$Ce$_x$CuO$_4$ and
  La$_{2-x}$Sr$_x$CuO$_4$ and found that the position of the Fermi
  level with respect to (the maximum of) the valence band was essentially
  constant, and approximately in the middle of the gap.
  The data of Allen strongly conflict with the `semiconductor model',
  because there one would expect that for electron or hole doped
  samples, the chemical potential is at the bottom of the conduction band
  or at the top of the valence band respectively
  [Fig.~\ref{fig:mux.models}c]. Thus, the difference in
  position of $\mu$ should be of the size of the gap ({\it i.e.} 1.5--2~eV
  \cite{Nucker87,Etemad88,Uchida89}), which is not found in the
  measurements \cite{Allen90}. Recent measurements by the same group
  on higher quality Nd$_{2-x}$Ce$_x$CuO$_4$ samples covering the whole
  doping range confirm this result \cite{Anderson93}.
  The chemical potential is constant as a function of doping, and the
  density of states at $\mu$ increases with doping by an apparent filling
  of the gap with states.

On the other hand, the `semiconductor model' is supported by
  measurements of van Veenendaal {\it et al.} \cite{Veenendaal93} in the
  Bi$_2$Sr$_2$Ca$_{1-x}$Y$_x$Cu$_2$O$_{8+\delta}$ system.
  Here, the transition from insulator to superconductor is made by
  varying the Y content from 1 to 0. According to van Veenendaal's
  data, this transition is accompanied with a rather strong shift
  in chemical potential \cite{Veenendaal93}.
  They find the same behaviour in the core
  level positions as well as in X-ray and UV-light excited valence
  band spectra, which adds confidence to their results.
  In the metallic regime, the changes in chemical potential are small,
  and can be described with the doping of a rigid narrow band
  \cite{Veenendaal93}. Such a small shift in the metallic regime was
  also found by Shen {\it et al.} who varied the doping in
  Bi$_2$Sr$_2$CaCu$_2$O$_{8+\delta}$ by changing the oxygen content
  \cite{Shen91}.
  Prior to the study of van Veenendaal {\it et al.}, at least two other
  groups have studied the Bi$_2$Sr$_2$Ca$_{1-x}$Y$_x$Cu$_2$O$_{8+\delta}$
  system with photoelectron spectroscopy.
  Itti {\it et al.} \cite{Itti91} measured all core level positions
  over the whole doping range and essentially found the same results
  as van Veenendaal {\it et al.} \cite{BSCYCO-doping-plot}.
  However, they do not consider the possibility of changes in chemical
  potential and instead try to explain their data fully in terms of
  chemical shifts.
  Kusunoki {\it et al.} \cite{Kusunoki91} limited themselves to the valence
  bands, from which they could defer only a small shift in chemical potential
  over the total Y doping range of $x=0$--0.6.

In this paper we present the results of a photoelectron study of
  La$_{2-x}$Sr$_x$CuO$_4$ pellets and thin films, in the Sr doping range
  $x=0$--0.25. This range covers both the parent undoped insulating
  material and the doped metallic compounds.
  The choice for the La$_{2-x}$Sr$_x$CuO$_4$ system was stimulated by the
  facts that it has a relatively simple crystal structure with only one
  CuO$_2$ plane per unit cell, and that it is hole doped with the doping
  level rather well defined by the Sr content.
  Note that the difference between the data on Nd$_{2-x}$Ce$_x$CuO$_4$
  and Bi$_2$Sr$_2$CaCu$_2$O$_{8+\delta}$ in principle could be caused
  by the unusual electron doping in the former compound.

The structure of the rest of the article is as follows.
  First we will describe the preparation and characterization of the
  samples used in this study, especially of the thin films. Then
  the XPS core level spectra will be quite extensively discussed, since
  not many photoelectron spectra of the La$_{2-x}$Sr$_x$CuO$_4$ system
  have been published in the literature until now.
  Finally the results of the behaviour of the chemical potential, as
  deduced from the shift in the photoelectron spectra, are presented
  and discussed.

\section{Preparation and characterization}
\label{sec:mux.char}

Starting materials for the preparation of the ceramic pellets are
  La$_2$O$_3$, CuO and SrCO$_3$ powders. The purity of the powders
  is better than 99.99~\%, except for SrCO$_3$ which is 99.5~\%
  pure with main contaminants Ca and Ba.
  The procedure followed in making the pellets is similar to that
  of Blank {\it et al.} \cite{Blank88,Blank91} for the preparation of
  YBa$_2$Cu$_3$O$_{7-\delta}$ and shortly is as follows.

Carefully weighted amounts of the oxides are dissolved in nitric acid
  and mixed with citric acid monohydrate. After the addition of
  ammonium hydroxide to make the solution pH-neutral (pH value
  approximately 6.8), it is heated to 300~$^{\circ}$C until a solid state
  reaction occurs where a very fine powder is formed. The advantage of this
  technique compared with conventional mixing and grinding of the
  powders is a very homogeneous distribution of the elements in the
  final powder since the cations were mixed as ions in a solution.
  After the solid state reaction, the powder is ground and annealed
  for 9 hours at 920~$^{\circ}$C in flowing oxygen or air. Subsequently,
  the powder is reground and pressed under 15 tons into pellets
  with a diameter of 15~mm. The final preparation step consists
  of sintering at 1150~$^{\circ}$C in air for 5~h and an extra postanneal
  at 750~$^{\circ}$C in 1~bar oxygen during 11~hours. In all cases, the
  cooling rate of the furnace is set to 1.5~$^{\circ}$C/min.

With this procedure, several samples of La$_{2-x}$Sr$_x$CuO$_4$ were made
with steps in
  Sr doping of 0.05, covering the total range $x=0$--0.25.
  It was not possible to prepare a pellet with $x=0.30$, since a
  suspension is formed in the acid solution after adding the
  ammonium hydroxide. This resulted in inhomogeneous samples.

After completion of all characterization and XPS measurements on this
  series of samples, the pellets were used as target in the laser
  ablation chamber for the in situ growth of thin films.
  Since the pioneering work of Moorjani {\it et al.} \cite{Moorjani87},
  very little work has been published about growth of La$_{2-x}$Sr$_x$CuO$_4$
  thin films with laser ablation. Only recently two papers have appeared on
  this subject. Yi {\it et al.} ablated from $x=0.15$ material and found
  a maximum $T_{\rm c}$ of 22~K for films grown at $\sim 825$~$^{\circ}$C
  in a 250~mTorr oxygen atmosphere \cite{Yi92}. Chern {\it et al.} deposited
  La$_{2-x}$Sr$_x$CuO$_4$ films using a combination of pulsed molecular
  oxygen and a continuous source of atomic oxygen, with a background pressure
  as low as 1~mTorr. For $x=0.15$ material grown at 700~$^{\circ}$C, they
  find a maximum $T_{\rm c}$ of 15~K \cite{Chern92}.
  The earliest in situ growth of high quality La$_{2-x}$Sr$_x$CuO$_4$
  films was done
  by Suzuki using sputtering techniques \cite{Suzuki87}. His films
  were well oriented, but also had lower critical temperatures with
  a maximum of 23~K. In a later study of the same author this was
  improved to 29~K \cite{Suzuki89}.
  The highest $T_{\rm c}$ so far of La$_{2-x}$Sr$_x$CuO$_4$ thin films is
  approximately 35~K,
  obtained by Kao {\it et al.} for 8000~\AA\ thick films \cite{Kao91}.
  According to all publications, the optimal growth conditions for
  in situ growth of La$_{2-x}$Sr$_x$CuO$_4$ thin films are similar to
  those of YBa$_2$Cu$_3$O$_{7-\delta}$.
  Thus, for the preparation of La$_{2-x}$Sr$_x$CuO$_4$ films in our
  YBa$_2$Cu$_3$O$_{7-\delta}$
  thin film growth.
  The temperature of the SrTiO$_3$ substrate during growth is
  750~$^{\circ}$C
  and the oxygen pressure is 750~mTorr. Assuming an equal ablation rate
  for YBa$_2$Cu$_3$O$_{7-\delta}$ and La$_{2-x}$Sr$_x$CuO$_4$,
  we estimate that the films are approximately
  250~nm thick.
  After ablation of the film, the chamber is filled with 800~Torr
  oxygen and subsequently  the sample is cooled in
  approximately 30~minutes in two steps to below 200~$^{\circ}$C.

The characterization of the pellets and thin films started with
  X-ray diffraction for determination of the structure and cell
  parameters, and detection of possible secondary phases in the
  samples. Temperature dependent resistance measurements revealed
  the (super)conducting properties and X-ray Photoelectron Spectroscopy
  (XPS) and Electron Probe Micro Analysis (EPMA)
  are used for compositional analysis.
  The results of this characterization will be presented and discussed
  here; the use of XPS for determination of the surface quality and
  shift in chemical potential versus doping is described in the next
  paragraph.

A typical result of a X-ray diffraction measurement of pellets and
  thin films is given in Fig.~\ref{fig:mux.lscoxrd}.
  The scans given in the figure are for samples with $x=0.15$
  and are typical for all samples.
  All peaks in the spectrum of the pellet can be assigned to the
  superconductor, and the majority of them are labeled in
  Fig.~\ref{fig:mux.lscoxrd}a with their Miller indices $(hkl)$.
  For other doping levels,
  sometimes a faint reflection was found at $2\theta = 32.90^\circ$,
  visible as a small shoulder on the $(110)$ peak. This tentatively
  was identified as the (020) diffraction peak \cite{Panson87} but
  also has been reported as due to the secondary phase material
  La$_{2-y}$Sr$_y$Cu$_2$O$_5$ \cite{Payzant90}.
  As far as can be judged from the diffraction scans of the films, also
  these are free of secondary phases.
  The presence of sharp $(00l)$ peaks indicates well-defined
  $c$-axis oriented material.
  Some films with non-zero doping have a small amount of $a$-axis
  grains.
  One film with $x=0.15$, grown under slightly different conditions,
  had detoriated surface and transport properties and also had
  a small peak in the diffraction scan at $2\theta=44.20^\circ$,
  probably from not fully oxidized copper.
  The results obtained on this sample are excluded from the present
  study.

Figure~\ref{fig:mux.lscoacx} gives the room temperature lattice
  parameters deduced from the $\theta$--2$\theta$ scans, as a
  function of the Sr content $x$ for both pellets and films.
  For comparison, also the values determined by Tagaki {\it et al.} are
  given \cite{Tagaki89}. Clearly, the lattice parameters of our
  pellets are in good agreement with these values.
  For the films, only the $c$-axis value can be determined from the
  scans. Similar to the trend seen in the data for the pellets, the
  $c$-axis increases for higher doping levels, but compared to the
  pellets the effect is significantly smaller. This can not be
  caused by an oxygen deficiency of the films, since extra anneals
  of the films with $x=0.15$ and $x=0.20$ in 1~bar oxygen at
  450~$^{\circ}$C and 900~$^{\circ}$C respectively,
  hardly gave rise to larger
  values (open circles in Fig.~\ref{fig:mux.lscoacx}b).
  The shorter $c$-axis is probably due to strain in the films.
  The in-plane lattice parameters of the films are $\sim 3$~\%
  smaller than those of the SrTiO$_3$ substrates. The resulting in-plane
  strain causes a compressive strain in the perpendicular $c$-axis
  direction, resulting in a contraction of the $c$-axis. A similar
  relation between in-plain strain and $c$-axis length was found in
  multilayers of Nd$_{1.83}$Ce$_{0.17}$CuO$_x$ and YBa$_2$Cu$_3$O$_{7-\delta}$
  \cite{Gupta90}.

In Figure~\ref{fig:mux.lscort} we show the resistance curves of the
  doped samples, again for pellets and films.
  The curves were measured with a standard four-point measurement
  setup. Contact was made to the samples by pressing four gold-plated
  brass pins onto the samples. Since the contact geometry was different
  for each sample, the absolute values of the resistance given in
  Fig.~\ref{fig:mux.lscort} can not be compared with each other.
  However, the trend (visible for pellets and films) that the
  resistance decreases for higher Sr doping is still significant.
  The zero-doped samples all showed semiconducting behaviour of the
  resistance, without any sign of superconductivity above 4~K.

A phase diagram of the dependence of $T_{\rm c}$ on the Sr doping is given
  in Figure~\ref{fig:mux.lscotcx}.
  Shown are the values of $T_{{\rm c},0}$, defined as the temperature
  where the resistance is 10~\% of the value at the onset. Data of
  Tagaki {\it et al.} \cite{Tagaki89} are again included for comparison.
  The solid line in the figure is a guide to the eye through these
  data.
  For the pellets the values of $T_{\rm c,0}$ closely follow this
  line. However, the onset of superconductivity in the $R(T)$ curves
  is almost constant (dashed line in Fig.~\ref{fig:mux.lscotcx}).
  Similar behaviour of the onset temperature was found in magnetization
  data in a later study of Tagaki {\it et al.} on high quality samples
  in the overdoped region ($x>0.15$) \cite{Tagaki92}.

The $T_{\rm c}$ values of the films are rather low and almost independent
  of Sr doping. The shape of the resistance curve near $T_{\rm c}$ is quite
  comparable for all films: the superconducting onset is near 16~K
  and the total width of the transition is approximately 5~K
  [Fig.~\ref{fig:mux.lscort}b]. As already mentioned above, all data
  in the literature to date show reduced $T_{\rm c}$ values for in situ grown
  thin films, ranging from 15 to 35~K \cite{Chern92,Suzuki87,Kao91}.
  Kao {\it et al.} have studied the dependence of $T_{\rm c}$ on the
  thickness of
  the films and found that on (100) SrTiO$_3$ substrates the $T_{\rm c}$
  increased
  from approximately 20 to well above 30~K if the film thickness was
  increased from 2000 to 8000~\AA\ \cite{Kao91}.
  They suggest that the low $T_{\rm c}$'s for the thinner films are caused
  by strain. Indeed, combining known values of $dT_{\rm c}/dp$, the
  compressibility of the lattice, and the lattice mismatch 
  between substrate and film, we find a lowering of the critical
  temperature with at least 10~Kelvin, not even taking into account
  the anisotropy in $dT_{\rm c}/dp$, as {\it e.g.} found in
  YBa$_2$Cu$_3$O$_{7-\delta}$ \cite{Welp92}.
  Additional evidence that strain indeed plays an important role
  lies in the fact that (110)/(103) oriented films have higher $T_{\rm c}$'s
  than films with (001) orientation \cite{Kao91}. In the former case,
  the planes of the superconducting material are not parallel to
  the substrate so that the strain in these planes is reduced.
  Another reason for the lower critical temperatures was proposed by
  Suzuki, who assigned it to inhomogeneity of the Sr distribution
  \cite{Suzuki89}. Such an inhomogeneity always exists to a certain
  extent in a solid solution and may be intrinsic to single
  crystalline La$_{2-x}$Sr$_x$CuO$_4$---also early bulk single
  crystals had lower $T_{\rm c}$'s
  \cite{Hidaka87}. In addition, the double peak structure in the Sr
  core level spectra [see Fig.~\ref{fig:mux.xpssr}] seems to indicate
  that strontium occupies two chemically different sites.
  Finally, the lower $T_{\rm c}$ values could be related to an oxygen
  deficiency of the films.
  To test for this possibility, we have given the samples
  with $x=0.15$ and 0.20 an extra anneal in pure oxygen for 12 hours
  at 500 and 900~$^{\circ}$C respectively. After these anneals,
  the $T_{\rm c}$
  was substantially higher, but still full superconductivity was not
  obtained above 20~K. Since the anneals are expected to
  be very effective in removing any oxygen deficiency, we do not
  think it plays a major role in causing the lower $T_{\rm c}$'s.

The composition of the samples is checked with XPS and EPMA. For the
  pellets, EPMA showed that the cation stoichiometry was within 5~\%
  equal to the starting composition in the preparation. With a slightly
  larger inaccuracy this is also true for the La/Cu ratio in the thin
  films. It is difficult to estimate the Sr content of the films with
  EPMA, since with the present thickness of the films (250~nm) also
  part of the SrTiO$_3$ substrate is probed.

Very similar results were obtained in XPS. Using the data in
  Fig.~\ref{fig:mux.xpsp} and \ref{fig:mux.xpsf} (further discussed
  in the next section) and calculated values for the photoelectron
  cross-section $\sigma$ \cite{Scofield76}, furthermore assuming
  a homogeneous distribution of the elements in the samples, the
  composition of the pellets and the films is within the $\sim
  15$~\% accuracy of such an analysis equal to the desired ratios.
  The La and Sr intensities both follow the doping trend very well.
  In the case of the films, the assumption of a random distribution
  of the elements is not justified given the layered nature of the
  material and the $c$-axis orientation of the films. Analogous to
  an earlier study \cite{Rietveld93}, we will therefore discuss the
  influence of the layeredness of the La$_{2-x}$Sr$_x$CuO$_4$ material
  on the intensities in the XPS spectra.
  There are three posibilities for forming an ideal surface of
  $c$-axis oriented La$_{2-x}$Sr$_x$CuO$_4$ depicted as model T1, T2,
  and T3 in Figure~\ref{fig:mux.lscoterm}.
  Since the model assuming a random distribution of the elements describes
  the relative intensities of the photoelectron peaks very well,
  it is directly clear that termination T2 fits the data best.
  For model T1 or T3 the relative La intensity would be too low
  or too high respectively. Also this intensity would further
  decrease or increase for larger exit angles of the photoelectrons,
  whereas in our measurements (not shown here) the relative intensities
  hardly show any change for varying exit angles.
  Of course, it is possible that in reality a combination of several
  terminations is present. For example, if the surface has equal
  amounts of terminations T1 and T3, this will give relative XPS
  intensities very similar to those of model T2.

\section{Photoelectron Spectra}

The pellets are attached with Ag-paint \cite{Ag-paint} to stainless
  steel sample holders, suited for transportation in our ultra
  high vacuum
  (UHV) system. The as-prepared samples have highly contaminated
  surfaces, that are cleaned by scraping with a diamond file.
  This is done in the sample preparation chamber attached to the
  UHV transport chamber. The background pressure in this chamber
  during scraping is below $2\times 10^{-8}$~mbar.
  Several cycles of scraping---in total removing several hundred
  $\mu$m of material---are needed to obtain clean surfaces, as judged
  by a low C~$1s$ peak and a small high binding energy shoulder in the O~$1s$
  spectrum. The sample with Sr doping $x=0.05$ was very hard and
  could not be cleaned sufficiently by scraping. Therefore, data
  of this sample have been omitted in this study.
  Apart from scraping, we also have tried to clean the surfaces with
  ion etching. This always resulted in damage of the surfaces, almost
  independent of the ions used (Ar, Ne, O).
  Annealing the samples in the MBE system in an oxygen or ozone
  atmosphere is very effective in removing carbon contaminations
  but produces unwanted oxides at the surface.
  Thus to us, scraping seems the best method for preparation of clean
  surfaces of ceramic La$_{2-x}$Sr$_x$CuO$_4$ pellets.

The in situ growth and transport of the La$_{2-x}$Sr$_x$CuO$_4$
thin films makes any
  further surface preparation of the films superfluous.
  Immediately after growth in the laser ablation chamber and cool
  down to below 200~$^{\circ}$C, the samples are removed from the heater
  and the ablation chamber is quickly pumped down to $10^{-5}$~mbar.
  Then, the films are transported via the UHV transfer chamber to
  the analysis chamber.
  The temperature at which the
  sample is removed from the heater block is critical.
  If sample transport is started at too high or too low temperatures,
  the samples are oxygen deficient or have contaminated surfaces
  respectively.

Within half an hour after the preparation of the pellets and the
  films a complete set of photoelectron spectra is recorded, consisting
  of the strongest core level peak of each element and the valence
  band.
  Then, this series of spectra is repeated for approximately 7~hours
  in order to obtain sufficient statistics.
  The pressure of the system with the X-ray source operating is in
  the low $10^{-10}$~mbar region; the base pressure is an order of
  magnitude lower.

During the whole experiment no change was visible in any of the
  spectra, except for the O~$1s$ spectrum. Due to the exposure of
  the surface to X-rays, possibly in combination with the low
  vacuum pressure, the main peak in this spectrum gradually shifted
  to higher binding energies [Fig.~\ref{fig:mux.xpsdam}].
  Since this is already visible in spectra taken 30~minutes after the
  first spectrum, all XPS measurements were started with the O~$1s$
  line. Only these initial spectra are given in
  Fig.~\ref{fig:mux.xpsp} and \ref{fig:mux.xpsf} and only these are
  used for the determination of shifts in peak position.
  From the shape of the O~$1s$ spectrum it is clear that the main
  O~$1s$ line consists of two components which probably originate
  from the two inequivalent oxygen sites in the crystal structure
  [Fig.~\ref{fig:mux.lscoterm}b].
  Apparently, due to the radiation, the low binding energy component decreases
  in intensity or shifts to higher binding energies causing an overall
  shift of
  the main oxygen line.
  Similar shifts of the main O~$1s$ line as a function
  of time were also found by Fowler {\it et al.} in their study of cleaved
  YBa$_2$Cu$_3$O$_{7-\delta}$ single crystals \cite{Fowler90}. In addition,
  they found an increase of the intensity of the high binding energy shoulder
  due to the high background pressure in their system ($\approx 1 \times
  10^{-9}$~Torr).
  In our case, this shoulder {\em decreased\/} in intensity, probably
  due to X-ray stimulated desorption of surface contaminants. Also
  the small C~$1s$ peak at 284~eV [Figs.~\ref{fig:mux.xpsp} and
  \ref{fig:mux.xpsf}] decreased in intensity after prolonged
  exposure of the surface to X-rays.

Other evidence for the influence of the X-ray radiation of the XPS
  source was found in room temperature work function measurements using our
  Kelvin probe setup. For the as-prepared pellets and films we on
  average find work function values of 4.7 and 5.6~eV respectively.
  After three minutes exposure to X-ray radiation from the XPS
  source, operated at a quarter of the normal power, the work function had
  changed by $\sim 30$~meV. At the end of the whole series of XPS
  measurements, the total change in work function typically was 50~meV.
  This indicates that most of the damage occurs in the early time
  of the X-ray exposure
  and then very rapidly saturates. The work function of the pellets increased,
  whereas that of the films always decreased.

An overview of all core level and valence band spectra of pellets
  and films is given in Figures~\ref{fig:mux.xpsp} and \ref{fig:mux.xpsf}
  respectively. For clarity, only the spectra for $x=0$, 0.10, and 0.25
  are shown and offsets are given to spectra of samples with non-zero
  doping.
  The spectra are measured using Mg~$K$$\alpha$ radiation, and
  the pass energy of the hemispherical analyzer is set to 20~eV,
  giving an instrumental broadening of 0.4~eV.
  The energy scale of the spectrometer is calibrated using a
  freshly evaporated Cu film, and published values of the binding energies
  of Cu core levels \cite{Briggs90}.

\subsection{La~$3d_{5/2}$ Spectrum}

The La~$3d_{5/2}$ spectrum consists of a double peak. For the undoped
  pellet, the main peak lies at 833.4~eV and a satellite is located
  at 4.40~eV higher binding energy. The splitting between the two
  features is independent of Sr doping.
  The separations found here are comparable with those in the
  La~$3d_{5/2}$ spectra of La$_2$O$_3$ \cite{Wagner79} and LaMO$_3$
  (M=Fe, Co, Al) \cite{Burroughs76}.
  The much smaller value reported by Viswanathan for La$_2$CuO$_4$ (3.1~eV,
  \cite{Viswanathan86}) and the larger value found in LaBaCuO by
  Steiner {\it et al.} (5.3~eV, \cite{Steiner87b}) are probably caused
  by the low quality of their samples.

Fuggle {\it et al.} have shown that the core level line shapes of, among
  others, the
  oxides of the early lanthanides can be explained as composed of a
  well screened peak at low binding energy and a poorly screened peak at
  higher binding energy \cite{Fuggle80}. Screening arises from the coupling of
  (partially) empty $4f$ levels with the delocalized occupied O~$2p$
  states. For La$_2$O$_3$, the initially empty $4f^1$ screening level
  is lowered in the final state of the photoelectron proces to below
  the Fermi level. As a consequence, the total energy of the final
  state can be lowered if an electron is transferred form the occupied
  oxygen states to this $4f^1$ level \cite{Schonhammer77}.
  In general, the role played by such empty screening levels strongly
  depends on the initial position above $E_{\rm F}$ and the coupling to the
  occupied levels \cite{Fuggle80}. The latter is directly reflected
  in the amplitude of the well screened peak. The former can be
  measured with inverse photoelectron spectroscopy (IPES). Data on
  La$_{2-x}$Sr$_x$CuO$_4$, {\it e.g.} those published
  by Gao {\it et al.} \cite{Gao87} and
  Riesterer {\it et al.} \cite{Riesterer87},  
  indeed show unoccupied La states close to $E_{\rm F}$.

In comparing the La~$3d_{5/2}$ spectra for films and pellets, several
  small differences become visible. Both the relative intensity and the
  splitting of the two peaks in the spectra of the pellets greatly
  resemble those of La$_2$O$_3$. For the films, the splitting is 0.2~eV
  reduced and the intensity of the unscreened peak is somewhat larger.
  Following the above explanation of Fuggle {\it et al.}, this could reflect
  a smaller coupling of the $4f$ levels to the valence band.
  Most likely, the smaller coupling is due to oxygen defects, or an
  elongation of the La-O bond in the film because of the lattice
  mismatch between the SrTiO$_3$ substrate and the film. Also Sr disorder
  (see below) might play a role.

\subsection{Cu~$2p_{3/2}$ Spectrum}

Just as in all other cuprate superconductors, the Cu~$2p_{3/2}$
  spectrum closely resembles that of CuO, consisting of a main peak with
    $|\underline{2p}_{3/2}3d^{10}\underline{L}\rangle$
  character and a
    $|\underline{2p}_{3/2}3d^9L\rangle$
  satellite peak. Here $\underline{L}$ denotes a hole on the
  neighbouring ligand oxygen of copper.
  The relative intensity of the satellite and main peak does not change
  as a function of doping and is equal to $\sim 0.36$, both for pellets
  and films. Only the pellet with $x=0.25$ has a somewhat lower
  satellite intensity, 0.31 of the main line. This can be due to a
  lower oxygen content, since samples with doping levels $x>0.15$ tend
  to be oxygen deficient \cite{Shafer87,Torrance88}.

Accurate determination of the peak position---and shifts therein---is
  difficult, as the Cu~$2p_{3/2}$ spectrum is rather broad. Therefore,
  the leading edge of the spectrum will be taken as reference for
  determining shifts in the peak position as a function of doping.
  This will give the same results, assumed that the
  spectra do not change shape as a function of doping. Inspection of
  Figs.~\ref{fig:mux.xpsp} and \ref{fig:mux.xpsf} shows that this is
  indeed the case.
  Note that van Veenendaal {\it et al.} found a significant narrowing of
  the Cu~$2p_{3/2}$ main line with increased Y doping in
  Bi$_2$Sr$_2$Ca$_{1-x}$Y$_x$Cu$_2$O$_{8+\delta}$.

\subsection{C~1s Spectrum}

The amount of carbon contamination of the pellets and the films is low.
  The spectra have been recorded with a higher pass energy of the
  electron analyzer (50~eV, total instrumental resolution $\sim 1$~eV),
  for higher sensitivity.
  Both pellets and films show a peak at 284~eV, the binding energy for carbon
  and carbonhydrydes, and in addition the pellets have a peak around
  289~eV, the binding energy of carbonoxides.
  The pellets have been scraped until this residual level of
  contamination was found. Spectra for $x=0.15$ and $x=0.20$ (not
  given in Fig.~\ref{fig:mux.xpsp}) have a lower intensity at 289~eV,
  but still it is not negligible \cite{Glastra93}.
  This contamination is either due to impurity material between the
  grains of the pellets, or a residual amount of carbonate used as
  one of the starting materials in the preparation. The surface of
  the films is probably contaminated during the cool down after
  preparation or during the UHV transport of the sample.
  The increase in intensity below 282~eV for the doped samples is the
  high binding energy tail of the Sr~$3p_{1/2}$ peak at 278~eV.

\subsection{O~$1s$ Spectrum}

The O~$1s$ spectrum is recorded using Al~$K$$\alpha$ radiation, because
  with the use of Mg~$K$$\alpha$ radiation the O~$1s$ line is on the
  background of La $MNN$ Auger peaks. The O~$1s$ spectrum is
  dominated by a large peak at 528.7~eV (position for $x=0$) and a
  shoulder around 531~eV. Compared to the main peak, the shoulder
  is lowest for the undoped samples.
  For the pellets, there is a clear correlation between the carbon peak
  at 289~eV and the intensity of the high binding energy shoulder in the
  O~$1s$ spectrum, but certainly not the entire shoulder originates from
  carbonates (see e.g.\ the spectra of the films).
  At present, no `shoulder-free' O~1s spectrum has been published for
  the La$_{2-x}$Sr$_x$CuO$_4$ compounds. Probably the best
  results have been obtained by Takahashi {\it et al.} in a combined
  XPS/UPS study of single crystals \cite{Takahashi88}.
  In their spectrum for undoped samples only an asymmetric tailing of
  the main line to higher binding energies is visible, comparable to our
  spectrum of the undoped film. For $x=0.08$, they also find a shoulder
  in the spectra, which they assign to detoriation of the surface. This
  detoriation occured very rapidly at room temperature.
  As already discussed above, we do not find an increase of the
  shoulder in our spectra over the 8~hours duration of our
  measurements---at most a small {\em decrease\/} is found
  [Fig.~\ref{fig:mux.xpsdam}], probably due to X-ray stimulated desorption
  of contaminants from the surface.
  Angle resolved measurements confirm that the shoulder in the O~$1s$
  spectrum is at least partly due to surface contamination.

\subsection{Valence Band}

The valence band spectra of the samples are broad and almost
  featureless, due to the low resolution of the measurements.
  For the same reason it is nearly impossible to detect the
  very small increase of the density of states near $E_{\rm F}$ with doping.
  In the figures, the spectra are scaled to the total intensity
  in the valence band after background subtraction.

In spite of the low resolution, a clear difference is visible
  between the spectra of the pellets and the films. All films
  give rectangularly shaped spectra, whereas for the pellets the
  valence band is more rounded.
  For the pellets, the spectra greatly resemble those of others
  \cite{Nucker87,Steiner87b,Shen87}. Bandstructure calculations
  indicate that the low binding energy part of the valence band has more
  Cu weight and that the O~$2p$ contribution is at higher binding energies
  \cite{Mattheis87,Pickett87}. The films seem to have extra
  intensity at around 2 and 5~eV, that thus originates from
  Cu and O respectively.
  Especially, the increase of intensity at low binding energies
  is strong, indicating a significant difference in Cu~$3d$
  states between pellets and films. A possible explanation
  of this difference
  can be the presence of impurities in the pellets, {\it e.g.}
  oxygen deficiency at the surface, since it is known that
  impurities cause a smearing of the features in the valence
  band of the host material.

\subsection{Sr~$3d$ Spectrum}

The most puzzling core level spectrum is the Sr~$3d$ spectrum.
  With doping, La ions are replaced by Sr, and since there is only
  one crystallographic site for La in the La$_{2-x}$Sr$_x$CuO$_4$ lattice
  [Fig.~\ref{fig:mux.lscoterm}b] one would expect a single doublet
  in the Sr~$3d$ spectrum. Instead, we find two doublets in this
  spectrum, both for pellets and thin films. To our best knowledge,
  only one spectrum of Sr in La$_{2-x}$Sr$_x$CuO$_4$ is published
  in the literature,
  in an early study of Steiner {\it et al.} \cite{Steiner87a}. They also
  find double peaks in the Sr~$3d$ spectrum, which might be due to
  low quality of their material.
  Other references for Sr spectra come from the Bi cuprate
  superconductors \cite{Hill88,Hillebrecht89}, SrO
  \cite{Doveren80,Young85} and SrTiO$_3$ \cite{Nagarkar91}.
  For the superconductors always a second doublet is found, with an
  intensity depending on the quality of the samples and the cleanliness
  of the surfaces. The best samples seem to have a single doublet in
  their Sr~$3d$ spectrum. However, the intensity ratio of the
  $3d_{3/2}$ and $3d_{5/2}$ peak in such spectra always exceeds 0.67,
  the value expected from the multiplicity of the two levels and which
  is also found in theoretical calculations of the relative
  cross-sections \cite{Scofield76}. Experimentally, this ratio was
  found by van~Doveren and Verhoeven in spectra of SrO \cite{Doveren80}
  and in SrTiO$_3$ by Nagarkar {\it et al.} \cite{Nagarkar91}.

In Figure~\ref{fig:mux.xpssr} we give a compilation of Sr~$3d$ spectra
  of several Sr containing conducting oxides. The lower spectrum is
  that of Sr in SrTiO$_3$ which clearly consists of a single doublet,
  and is shifted by -1.6~eV in order to put the chemical potential of
  the sample in the middle of the gap \cite{Rietveld93}.
  Assuming gaussian line shapes, the spectrum can be well
  fitted using an energy splitting in the doublet of 1.75~eV and a
  relative intensity of 0.70 (solid line in the figure). The second
  curve is of a single crystal of Bi$_2$Sr$_2$CaCu$_2$O$_{8+\delta}$,
  in situ cleaved in the spectrometer chamber. This spectrum is entirely
  typical for the best spectra obtained for this material,
  {\it e.g.} those of Hill {\it et al.} and Hillebrecht {\it et al.}
  \cite{Hill88,Hillebrecht89}.
  The top two curves are for the La$_{2-x}$Sr$_x$CuO$_4$ thin films and
  pellets
  ($x=0.15$). Here, the background in the spectrum (probably
  originating from an energy-loss tail of the La~$4d$ core level
  at 103~eV) was removed by subtraction of the spectrum
  for zero doping [Figs.~\ref{fig:mux.xpsp} and \ref{fig:mux.xpsf}].
  For SrTiO$_3$ and Bi$_2$Sr$_2$CaCu$_2$O$_{8+\delta}$,
  the background has been removed using the
  Shirley method \cite{Shirley72}.
  If we take the energy splitting and intensity ratio from the Sr~$3d$
  spectrum of SrTiO$_3$ as reference, those of the three superconducting
  cuprates can be fitted with two doublets (solid lines in
  Fig.~\ref{fig:mux.xpssr}). The low binding energy doublet lies at 132.2~eV,
  except for the La$_{2-x}$Sr$_x$CuO$_4$ thin film,
  where it is 0.5~eV higher;
  the
  high binding energy doublet has its $3d_{5/2}$ peak at 134.1~eV.
  The relative intensity of the
  high binding energy doublet is 0.12, 0.55, and 1.0 times the intensity
  of the
  low binding energy peaks for Bi$_2$Sr$_2$CaCu$_2$O$_{8+\delta}$,
  La$_{2-x}$Sr$_x$CuO$_4$ pellets and films respectively.
  Note that the shape of the spectra for the La$_{2-x}$Sr$_x$CuO$_4$
xp
  \ref{fig:mux.xpsf}].

An explanation of the double peak structure in the Sr~$3d$ spectra
  is difficult.
  It is known that in the Bi$_2$Sr$_2$CaCu$_2$O$_{8+\delta}$ material
  Sr--Ca disorder easily
  occurs. The second doublet in this case thus could be due to Sr
  located at Ca sites in the lattice. Alternatively, since the
  relative intensity of the second doublet is so low, the spectrum
  may be viewed as a single doublet with non-gaussian line shapes
  (tail to higher binding energies).
  For the La$_{2-x}$Sr$_x$CuO$_4$ samples, the intensity is too strong
  for such an
  explanation.
  In their study of La$_{2-x}$Sr$_x$CuO$_4$ pellets, Steiner
  {\it et al.} assign the second
  doublet to oxygen vacancies \cite{Steiner87a}. Since the intensity
  of this peak increases for larger exit angles of the photoelectrons,
  they assume that the density of the vacancies increases near the
  surface.
  For our samples, we only find a very small increase of the high
  binding energy
  part of the Sr~$3d$ spectrum if the exit angle is increased. A pure
  surface phase, {\it e.g.} SrO, may thus be excluded. As already discussed
  in relation to the structural and transport properties of the samples
  [Section~\ref{sec:mux.char}], we have tested for the possible
  presence of O vacancies by in situ annealing the $x=0.15$ film at
  450~$^{\circ}$C in oxygen. This hardly produced any difference in the
  Sr~$3d$ spectrum.
  Therefore, a more probable origin of the double peak for the
  La$_{2-x}$Sr$_x$CuO$_4$
  samples are inhomogeneities in the Sr doping.
  This inhomogeneity should occur on very small length scales since
  the X-ray diffraction measurements do not give any indication of such an
  inhomogeneity \cite{Tagaki92}. XPS, on the other hand, is sensitive
  to the local chemical environment of the atoms.
  Note that due to the small coherence length, also the superconducting
  properties are sensitive to local changes in crystal or electronic
  structure.
  Further investigations, preferably using other analysis
  techniques that are sensitive to the local chemical environment of
  the ions, are certainly needed to clarify this point.

\section{Doping Dependence of the Chemical Potential}

The aim of the present XPS study was to find systematic shifts in
  the spectra as a function of Sr doping, signature of a doping
  dependence of the chemical potential.
  Fig.~\ref{fig:mux.lscoshift} shows the measured shifts for the
  pellets and the films, relative to the samples with $x=0.15$.
  For the determination of the shifts, we have taken the peak position
  in the O~$1s$ and Sr~$3d$ spectra. In the case of the Cu~$2p$,
  La~$3d$, and valence band spectra, which are characterized by
  broad peaks, instead the leading edge of the spectrum has been
  used. 
  Even though differences exist between the two sets of data in
  Fig.~\ref{fig:mux.lscoshift}, the general trend is similar. In
  both cases, a significant shift is found if the doping is varied.
  Furthermore, this shift is smooth over the whole doping range.
  An important difference with the data of van Veenendaal {\it et al.}
  obtained
  on Bi$_2$Sr$_2$Ca$_{1-x}$Y$_x$Cu$_2$O$_{8+\delta}$, is that no
  large shift is found at the transition from
  the insulating, undoped ($x=0$) material to the (super)conducting,
  doped ($x \geq 0.10$) samples.
  In this context it is important to mention that we have checked
  for the possible effect of charging during the measurements. In
  all cases, also for the undoped samples, we do not find a measurable
  change of the peak positions if the intensity of the X-ray source
  is varied.

The scatter of the core level positions of the different elements
  around the average (solid line in Fig.~\ref{fig:mux.lscoshift}),
  is caused by chemical effects, such as changes in Madelung
  potential or effective charge upon doping.
  As an example of the latter, one would expect that the LaO layers
  become more negative with Sr doping, since the La$^{3+}$ ions are
  replaced by Sr$^{2+}$. This will result in a larger shift of the
  La and Sr spectra. The CuO$_2$ planes, on the other hand, are doped
  with holes if the Sr content is increased, and consequently the Cu
  spectra will shift less.
  On average the shift of the core level spectra will of course reflect
  the changes in chemical potential.

The fact that there is not much difference in chemical potential shift for
  the pellets and the films indicates that the lower $T_{\rm c}$ of the
  films, the difference in Sr and valence band spectra, and the different
  levels of surface contamination for pellets and films apparently
  do not play a large role.
  Concerning the influence of contamination, it is illustrative to note
  that van Veenendaal {\it et al.}, who claim to have very clean surfaces
  \cite{Veenendaal93}, also find almost the same shifts in the photoelectron
  spectra of Bi$_2$Sr$_2$Ca$_{1-x}$Y$_x$Cu$_2$O$_{8+\delta}$ as
  Itti {\it et al.}, whose surfaces clearly suffer from a non-negligible
  amount of contamination \cite{Itti91,Gopinath92}.

\section{Discussion}

The main result that can be deduced from the changes in the XPS spectra
  is that the chemical potential $\mu$ in La$_{2-x}$Sr$_x$CuO$_4$
  shifts as a function of doping, without any discontinuity at the
  transition from metallic to insulating samples.
  Strong evidence that the shifts in the photoelectron spectra
  indeed are due to changes in $\mu$ is provided by the great
  similarity of the average core level shift and the shift of
  the valence band (compare the solid and dashed lines in
  Fig.~\ref{fig:mux.lscoshift}a and b).

The behaviour of the chemical potential found here in
  La$_{2-x}$Sr$_x$CuO$_4$ is different from that in van Veenendaal's
  study of Bi$_2$Sr$_2$Ca$_{1-x}$Y$_x$Cu$_2$O$_{8+\delta}$ {\em and}
  from that in Allen's study of Nd$_{2-x}$Ce$_x$CuO$_4$.
  Van Veenendaal {\it et al.} found a strong shift in $\mu$ at the
  transition from metallic to insulating samples \cite{Veenendaal93},
  which is clearly absent in our data. Also Takahashi {\it et al.} do not
  find a strong shift in La$_{2-x}$Sr$_x$CuO$_4$, comparing spectra of
  $x=0$ and $x=0.08$ cleaved single crystals \cite{Takahashi88}.

Allen {\it et al.} report that in Nd$_{2-x}$Ce$_x$CuO$_4$, the chemical
  potential does not shift at all as a function of doping \cite{Allen90}.
  This conclusion is based on a valence band study
  of an undoped and optimally doped sample. Recently this result was
  confirmed by the same group, using more samples in a wider doping
  range \cite{Anderson93}.
  However, for La$_{2-x}$Sr$_x$CuO$_4$, we find that the spectra do shift
  if the doping is changed. The shift is quite comparable to that found by
  Shen {\it et al.} and van Veenendaal {\it et al.} in the metallic regime
  of Bi$_2$Sr$_2$Ca$_{1-x}$Y$_x$Cu$_2$O$_{8+\delta}$. Following their
  explanation, also in La$_{2-x}$Sr$_x$CuO$_4$ the behaviour of $\mu$ can
  thus be described by the doping of a rigid band.
  In recent bandstructure calculations calculations by Czyzyk and
  van der Marel of La$_2$CuO$_4$ \cite{Czyzyk92}, a shift of 0.15~eV is
  predicted, adopting a rigid band model, if the doping increases
  from $x=0$ to 0.25. This is quite similar to the trend found in our
  experimental data.

\section{Conclusions}

We have performed a systematic study of the behaviour of the chemical
  potential as a function of doping in the La$_{2-x}$Sr$_x$CuO$_4$
  system. For this study, both pellets
  and films were made with Sr doping $x$ ranging from 0 to 0.25.
  Extensive characterization of the samples shows that the pellets
  have good structural and electrical properties. The thin films
  are well-oriented but have reduced $T_{\rm c}$
  values. Similar low $T_{\rm c}$'s were found by others, who assigned
  it to strain in the layers. We found additional evidence for this
  assignment in the smaller increase of the $c$-axis length versus
  Sr doping of the films as compared to that of the pellets.

 From the relative peak intensities in the XPS measurements we have
  deduced that if the surface has only one type of surface termination
  this must be a LaO layer directly followed by a CuO plane.
  In the discussion of the separate XPS core level spectra, special
  attention was paid to the Sr~$3d$ spectra. In spite of the fact that
  the La$_{2-x}$Sr$_x$CuO$_4$ crystal structure has only one
  crystallographic site for Sr, we find two doublets in all Sr~$3d$
  spectra indicating Sr occupies two chemically different sites.
  We suppose that this is caused by local inhomogeneity of the
  Sr doping.

The shift that is found in all photoelectron spectra upon doping, shows
  that the chemical potential of La$_{2-x}$Sr$_x$CuO$_4$ continuously
  changes as a function of Sr content.
  No large shift in $\mu$ was found at the metal-insulator
  transition, in contrast with predictions of the `semiconductor model'
  of the doping behaviour of $\mu$ in high-$T_{\rm c}$ materials.
  On the other hand, our data also do not support the `impurity model'
  where it is supposed that the chemical potential is constant as a
  function of doping.
  The results on La$_{2-x}$Sr$_x$CuO$_4$ can be
  best explained assuming that the shift in $\mu$ is caused by the
  doping of a rigid narrow band, lying near the middle of the gap.

\section{Acknowledgements}

We thank L. Lander for technical assistance with the preparation of
  the thin films.
  This investigation was supported by the Netherlands Foundation for
  Fundamental Research on Matter (FOM), and the Dutch National
  Research Program for high-$T_{\rm c}$ superconductivity (NOP).


  \figure{Schematic picture of the doping dependence of the
      chemical potential $\mu$ in the high-$T_{\rm c}$ superconductors.
      {\rm (a)} Insulating parent compound with $\mu$ located in
      the middle of the charge-transfer gap.
      {\rm (b)} Doped compound, according to the `impurity model'.
      The position of $\mu$ is constant and the gap is filled with
      doping induced states.
      {\rm (c)} Doped compound, according to the `semiconductor
      model'. The chemical potential $\mu$ has moved to the top of the
      valence band (VB) or to the bottom of the conduction band (CB),
      depending on the type of doping. Possible transfer of
      spectral weight caused by the doping proces has been
      neglected in the picture.
    \label{fig:mux.models}}

  \figure{X-ray diffraction scans of a La$_{2-x}$Sr$_x$CuO$_4$ pellet
      {\/\rm (a)} and a La$_{2-x}$Sr$_x$CuO$_4$ thin film {\/\rm (b)} made
      by laser ablation from this pellet.
      The doping level of the samples is $x=0.15$;
      the scans given here are typical for all doping levels.
      Cu $K$$\alpha$ radiation with a Ni filter is used as radiation
      source.
      All peaks in the scans can be assigned to the superconductor or,
      in the case of the film, to the substrate. Within the sensitivity
      of the measurements, no secondary phases are detected.
      The majority of the peaks are labeled with their corresponding
      $(hkl)$ Miller indices; substrate peaks are marked with S$(h00)$
      and X-ray satellites with an asterisk.
      Note the logarithmic intensity scale for the film {\/\rm (b)}.
    \label{fig:mux.lscoxrd}}

  \figure{Room temperature lattice constants $a$, $b$ {\/\rm (a)} and
      $c$ {\/\rm (b)} of La$_{2-x}$Sr$_x$CuO$_4$ samples as a function of
      Sr doping $x$. Values for pellets and films are denoted by filled
      squares and circles respectively.
      For comparison, values determined by Tagaki {\it et al.} are given
      as well (crosses) \protect\cite{Tagaki89}. The lattice
      parameters of the pellets agree very well with these data.
      The films have significantly smaller $c$-axes, that
      only slightly increase after extra oxygen anneals (open
      circles).
      Note that for the orthorhombic samples, doping level
      $x < 0.10$, normalized lattice parameters $a/\protect\sqrt{2}$,
      $b/\protect\sqrt{2}$ and $c$ are shown.
    \label{fig:mux.lscoacx}}

  \figure{Temperature dependent resistance of La$_{2-x}$Sr$_x$CuO$_4$
      pellets {\/\rm (a)} and thin films {\/\rm (b)} for different Sr
      dopings $x$.
      The films are prepared by laser ablation, where the pellets were
      used as target.
      Due to different contact geometries during the resistance
      measurements, absolute values of the resistance can not be well
      compared with each other. Still, the observed decrease in
      resistance for increased Sr doping is significant.
      Note the constant onset temperature of the superconducting
      phase transition in the pellets and the similarly constant
      (but reduced) $T_{\rm c}$ values of the films.
    \label{fig:mux.lscort}}

  \figure{Dependence of the superconducting phase transition temperature
      $T_{\rm c,0}$ of La$_{2-x}$Sr$_x$CuO$_4$ on the Sr doping $x$, for
      pellets (squares) and thin films (circles). Results of Tagaki
      {\it et al.} (crosses) are included for comparison
      \protect\cite{Tagaki89}.
      The solid curve is a guide to the eye through these data.
      The dashed line is the value of $T_{\rm c,onset}$ of our pellets,
      which is remarkably constant.
      Open circles give the $T_{\rm c}$'s of films after an extra anneal in
      oxygen.
    \label{fig:mux.lscotcx}}

  \figure{Models for forming an ideal surface of $c$-axis oriented
      La$_{2-x}$Sr$_x$CuO$_4$. In {\/\rm (a)} the three possible layer
      sequences T1--T3 are given (the first row is the top layer) for
      the La$_{2-x}$Sr$_x$CuO$_4$ crystal structure as schematically
      depicted in {\/\rm (b)} (from Hass \protect\cite{Hass89}).
    \label{fig:mux.lscoterm}}

  \figure{O~$1s$ core level of a La$_{2-x}$Sr$_x$CuO$_4$ thin film
      ($x=0.15$) as a function of X-ray exposure time.
      Solid line: 2~min, and dashed line: 420~min.
      The shift of the main peak is 0.15~eV. The decrease of the
      high binding energy shoulder with time points towards X-ray stimulated
      desorption of surface contaminants.
    \label{fig:mux.xpsdam}}

  \figure{Core level and valence band photoelectron spectra of
      La$_{2-x}$Sr$_x$CuO$_4$ pellets as a function of Sr doping $x$.
      For clarity, the spectra for non-zero doping have been given an
      offset.
      The carbon spectra have been taken with lower resolution for
      higher sensitivity. Futher discussion of the spectra is given
      in the text.
    \label{fig:mux.xpsp}}
  \figure{Core level and valence band photoelectron spectra of
      La$_{2-x}$Sr$_x$CuO$_4$ thin films as a function of Sr doping $x$.
      For clarity, the spectra for non-zero doping have been given an offset.
      The carbon spectra have been taken with lower resolution for
      higher sensitivity. Futher discussion of the spectra is given
      in the text.
    \label{fig:mux.xpsf}}

  \figure{Sr~$3d$ core level spectrum of several Sr containing oxides,
      recorded with a Mg~$K$$\alpha$ radiation source.
      From top to bottom (symbols):
      La$_{2-x}$Sr$_x$CuO$_4$ film, La$_{2-x}$Sr$_x$CuO$_4$ pellet,
      Bi$_2$Sr$_2$CaCu$_2$O$_{8+\delta}$ single crystal, and SrTiO$_3$
      single crystal. The curves have been given an offset
      for clarity; each curve is scaled such as to give equal
      intensity of the low binding energy component in the spectrum.
      The solid lines are fits to the spectra, assuming the presence
      of two chemically shifted doublets. The intensity of the high
      binding energy doublet is (from top to bottom) 1.0, 0.55, 0.12
      and 0 times
      that of the low binding energy one.
    \label{fig:mux.xpssr}}

  \figure{Core level and valence band position of
      La$_{2-x}$Sr$_x$CuO$_4$ pellets {\/\rm (a)} and thin films
      {\/\rm (b)} as a function of Sr content $x$. The positions
      are given relative to those of the $x=0.15$ samples.
      The dashed lines connect the positions of the valence band
      leading edge; the solid lines follow the average core level
      positions.
      Within a noise band of 0.2~eV, a clear shift is found in the
      spectra, indicating a doping dependence of the chemical potential.
    \label{fig:mux.lscoshift}}

\end{document}